\documentclass[aps,prl,amssymb,showpacs,twocolumn]{revtex4}
\usepackage{graphicx}
\usepackage{amsmath,bm}
\usepackage[mathscr]{euscript}

\begin{document}
\title{Structural Evolution of 1D Spectral Function from Low- to High-Energy 
Limits}

\author{Hideaki Maebashi and Yasutami Takada}
\affiliation{Institute for Solid State Physics, University of Tokyo, Kashiwa, 
Chiba 277-8581, Japan}


\begin{abstract}
By exactly analyzing the spin-$\frac{1}{2}$ Luttinger liquid (LL) and numerically 
solving a model of a mobile impurity electron in the LL, we obtain the 
one-electron spectral function $A(p,\omega)$ in a one-dimensional (1D) metal in 
an entire range of $p$ at zero temperature. For $|p|$ near the Fermi point 
$p_{\rm F}$, $A(p,\omega)$ is featured by two prominent peaks of spinon and 
(anti)holon representing spin-charge separation, but we also find an additional 
cusp structure between them. For $|p| \!\gg\! p_{\rm F}$, this structure evolves as 
a main peak in $A(p,\omega)$ by swallowing the antiholon mode and its dispersion 
relation approaches the one of a free electron, implying the existence of an 
electron excitation in the whole region, but not quite a quasiparticle in the 
Fermi liquid due to ever existing power-law decay of the excitation.
\end{abstract}

\maketitle
 

The concept of spin-charge separation plays a central role in describing 
low-energy physics near Fermi points in a 1D interacting electron gas, a typical 
example of the spin-$\frac{1}{2}$ Luttinger liquid (LL)~\cite{Tomonaga50}. This 
concept may be confirmed in real materials by various experiments~\cite{Experiment}, 
including the recent high-resolution angular resolved photoemission spectroscopy 
in which the one-electron spectral function $A(p,\omega)$ can be directly 
measured in the wide range of momentum $p$ and energy $\omega$.

If $|p|$ is not restricted to the region near the Fermi momentum $p_{\rm F}$, 
the linear spectrum approximation, usually adopted in the LL theory, is not 
sufficient in appropriately obtaining $A(p, \omega)$. In fact, the effect of the 
nonlinear spectrum on $A(p, \omega)$ has been intensively studied in recent 
years~\cite{ISG12,Penc95,IG09,PWA09,SIG10,Essler10}. According to those studies 
on integrable systems, $A(p,\omega)$ has singularities for arbitrary $p$ as 
$A(p,\omega) \propto |\omega - \epsilon_{\nu}(p)|^{-\mu_{\nu}(p)}$ with $\nu 
\!=\! s$ and $c$, where $\epsilon_{s}(p)$ and $\epsilon_{c}(p)$ are energies of 
spin and charge collective excitations, respectively. In the usual LL theory, 
the exponent $\mu_{\nu}(p)$ is independent of $p$, but the nonlinearity in the 
electron dispersion makes it depend on $p$~\cite{IG09,PWA09,SIG10,Essler10,Meden99}. Since the edge of 
support of $A(p,\omega)$ is located at $\omega\!=\!\epsilon_{s}(p)$, $\mu_{s}(p)$ 
determines the power of the threshold singularity in $A(p,\omega)$ and its actual 
value has been given from the finite-size spectrum obtained by the Bethe-ansatz 
method~\cite{PWA09,Essler10}. For nonintegrable systems, this threshold 
singularity remains intact, but the singularity at $\omega\!=\!\epsilon_{c}(p)$ 
is smeared into a broad peak~\cite{SIG10}. 

In those preceding works, only the singularities at $\omega=\epsilon_{s}(p)$ and 
$\epsilon_{c}(p)$ are discussed on the belief that the electron nature will 
not sustain in the spin-charge separated system. For $|p|$ far away from 
$p_{\rm F}$, however, the effect of interactions becomes so weak that we would 
naively expect that the nature of an injected electron to measure $A(p,\omega)$ 
manifests itself as a main peak in $A(p,\omega)$. Then a natural question 
arises: {\it  Does an electron-like excitation mode actually exist in the 1D 
interacting electron gas for $|p| \! \gg \! p_{\rm F}$?} If yes, a related and more 
intriguing question is: {\it How does the electron-like mode reconcile with 
the physics of spin-charge separation for $|p| \! \approx \! p_{\rm F}$?}

In this Letter, we have carefully studied the 1D one-electron Green's function 
$G(p,t)$ in momentum space and time and found that for $p \! \approx\! p_{\rm F}$, 
its long-time asymptotic form is composed of {\it three} independent modes of 
power-law decay. Two of them correspond to well-known spinon and (anti)holon 
excitations, but the rest describes the mode of an electron-like particle 
({\it pseudoelectron}) which may be regarded as an electron dressed with a 
``cloud'' of low-lying spin and charge collective excitations. This pseudoelectron 
does not appear as a main structure in $A(p,\omega)$ and never leads to a finite 
jump in the momentum distribution function $n(p)$, but it is considered as the 
1D counterpart of the Landau's quasiparticle in higher dimensions. As $p$ goes 
away from $p_{\rm F}$, the pseudoelectron structure gets broader, but with the 
further increase of $p$, it becomes less broad and eventually for $p \! \gg \! 
p_{\rm F}$, it evolves as a main and divergent peak in $A(p,\omega)$ 
by swallowing the antiholon mode. Concomitantly, its dispersion relation 
approaches the one of a free electron, allowing us to regard the pseudoelectron 
as a free electron, but actually it is not quite, nor the Landau's 
quasiparticle, basically because this excitation is accompanied by 
power-law decay. Those results clarify the generic feature of $A(p,\omega)$ 
in a 1D metal and answer the aforementioned two questions. In the following, 
we shall substantiate our claim. 

Let us consider the spin-$\frac{1}{2}$ Luttinger model, for which $G(p,t)$ with 
$p\!=\!p_{\rm F} \!+\! k$ is well known and is given by~\cite{Solyom79,Suzumura80}
\begin{align}
& 
{\rm i} G (p_{\rm F} \!+\! k,t) \!
= \! \int_{- \infty}^{\infty} \frac{{\rm d} x}{2 \pi} 
\frac{{\rm e}^{- {\rm i} k x}}{\eta \!- \!{\rm i}(x \! -\! v_{\rm F} t) } 
\frac{ 1 \!- \!{\rm i} \Lambda (x \!- \!v_{\rm F} t)}{
\sqrt{ 1 \!- \!{\rm i}  \Lambda (x \!-\! u_{s} t)} }
\nonumber
\\
& \hspace{1.3cm} \times
\left[ 1 \!-\! {\rm i}  \Lambda (x \!-\! u_{c}t) \right]^{-\frac{1 + \theta}{2}} 
\left[ 1 \!+\! {\rm i} \Lambda (x \!+\! u_{c}t)  \right]^{-\frac{\theta}{2}},
\label{eq:1}
\end{align}
for $t\!>\!0$ with $\eta$ a positive infinitesimal and $\Lambda$ a finite 
momentum-transfer cutoff. Here $u_s$ and $u_c$ are the velocities of spinon 
and (anti)holon, respectively, and $\theta$ denotes the power of singularity 
in $n(p)$. They are related to the velocity of an electron $v_{\rm F}$ and the 
interactions between electrons, $g_4$ and $g_2$, in the original Hamiltonian 
through $u_s \!=\!v_{\rm F}\!-\!g_{4}/2\pi$, $u_c\!=\![(v_{\rm F}\!+\!g_{4}/2
\pi)^2 \!-\!(g_2/\pi)^2]^{1/2}$, and $\theta \!=\!(v_{\rm F}\!+\!g_{4}/2\pi\!
-\!u_c)/2u_c$~\cite{comment1}. In a Galilean-invariant system, $v_{\rm F}$ is 
given by $v_{\rm F}\!=\!p_{\rm F}/m\!-\!g_{4}/2\pi \!+\!g_2/\pi$. By eliminating 
$g_{4}$ from those equations, we obtain $v_{\rm F}\!=\![(1\!+\!2\theta)u_c \!
+\! u_s]/2$, and thus $v_{\rm F}$ as well as $u_s$, $u_c$, and $\theta$ can be 
determined by the Bethe-ansatz method for integrable systems. 

By regarding the integrand in Eq.~(\ref{eq:1}) as an analytic function of $x$ 
to deform the integral path along the real axis into the lower-half complex $x$ 
plane and assuming $u_s < v_{\rm F} < u_c$, we can evaluate ${\rm i} 
G(p_{\rm F}+k,t)$ for $k>0$ in the long-time limit of $t \to \infty$ as 
\begin{align}
&{\rm i} G(p_{\rm F}\!+\!k,t) = 
e^{-k/\Lambda} 
\sum_{\nu = s, c} \left( \frac{\Lambda}{k} \right)^{1/2-\gamma_{\nu}} 
\frac{ {\rm e}^{- {\rm i} u_{\nu} k t + {\rm i} \pi \phi_{\nu}^{LL}/2}}{
( \Omega_{\nu}^{LL} t )^{1- \mu_{\nu}^{LL}}}
\nonumber
\\
& \hspace{1.7cm}
+ ( \Omega_{\rm F} t )^{- 1 + \mu_{\rm F}} 
{\rm e}^{- {\rm i} v_{\rm F} k t + {\rm i} \pi \phi_{\rm F}/2}. 
\label{eq:2}
\end{align}
The first two terms specified by $\nu\!=\!s$, $c$ represent the contributions from 
contours along the two branch cuts (which we take parallel to the imaginary axis) 
associated with the branch points at $x \!=\! u_s t \!-\! {\rm i}/\Lambda$ and 
$u_c t \!-\! {\rm i}/\Lambda$, showing the well-known time correlation specific 
to spinon and (anti)holon in the LL with the exponents, $\mu_{s}^{LL}$ and 
$\mu_{c}^{LL}$, given respectively as $\mu_{s}^{LL}\! =\! 1\! +\! \phi_{s}^{LL} 
\!=\! 1/2 \!-\! \theta$ and $\mu_{c}^{LL} \!= \!\phi_{c}^{LL}\! =\!(1\!-\! 
\theta)/2$. Here $\gamma_{s}\!=\!0$, $\gamma_{c}\!=\!\theta/2$, $\Omega_s^{LL}\!
=\! [(u_c \!+\!u_s)^{\theta/2}(u_c \!-\! u_s)^{(1\!+\!\theta)/2} \sqrt{\pi} ]^{2/
(1\!+\!2 \theta)}\Lambda$, and $\Omega_c^{LL}\!=\![(2 u_c)^{\theta/2}(u_c \!-\! 
u_s)^{1/2}\Gamma(\frac{1+\theta}{2})]^{2/(1 \!+\! \theta)} \Lambda$ with $\Gamma(z)$ 
being the Gamma function. The last term
stands for the contribution from a simple pole at $x \!=\! v_{\rm F}t \!-\!{\rm i}\eta$~\cite{SM93}, 
seemingly describing an electron moving with the velocity $v_{\rm F}$, but not 
quite a usual quasiparticle in the Fermi liquid because of the existence of the 
power-law decay with the exponent $\mu_{\rm F}\ (\neq \! 1)$, where $\Omega_{\rm F} \!=\! [(u_c \!-\! 
v_{\rm F})^{(1\!+\!\theta)/2}(u_c\!+\!v_{\rm F})^{\theta/2}(v_{\rm F}\!-\! 
u_s)^{1/2}]^{1/(1\!+\!\theta)} \Lambda$ and $\mu_{\rm F}\!=\! \phi_{\rm F}\!=\!-\theta$.
Thus we shall call it a 
{\it pseudoelectron}. Its decaying behavior persists even at $k \!\to\! 0$ 
and eliminates a finite jump in $n(p)$ at $p \!=\!p_{\rm F}$.

Casting ${\rm i}G(p,t)$ in Eq.~(\ref{eq:2}) into the form 
\begin{align}
&{\rm i} G (p,t) =   
\sum_{\ell = s, c, e}  [\Omega_{\ell}(p) t]^{- 1 + \mu_{\ell}(p)}
{\rm e}^{- {\rm i} \epsilon_{\ell}(p) t + {\rm i} \pi \phi_{\ell}(p)/2},
\label{eq:3}
\end{align}
and evaluating $A(p,\omega)$ by $\pi^{-1}{\rm Re} \int_{0}^{\infty}{\rm d}t 
{\rm e}^{{\rm i} \omega t}{\rm i} G(p,t)$ for $\omega \!>\! 0$, we can see 
that $A(p,\omega)$ has a singularity (peak or shoulder) at $\omega\!=\!
\epsilon_{\ell}(p)$ for $|\mu_{\ell}(p)| < 1$, in such a way as 
\begin{align}
A(p, \omega) = {\rm const.} + C_{\ell}^{\pm}(p) 
|\omega - \epsilon_{\ell}(p)|^{-\mu_{\ell}(p)},
\label{eq:4}
\end{align}
where $C_{\ell}^{\pm}(p)\!=\!\pi^{-1}\Omega_{\ell}(p)^{-1 + \mu_{\ell}(p)}
\Gamma\bigl (\mu_{\ell}(p)\bigr )\! \cos(\pi[\mu_{\ell}(p)\!\pm \! 
\phi_{\ell}(p)]/2)$ with the upper (lower) sign for $\omega$ larger (smaller) 
than $\epsilon_{\ell}(p)$. Since the edge of support of $A(p, \omega)$ is 
located at $\omega\!=\! \epsilon_{s}(p)$, $\phi_{s}(p)$ necessarily equals 
$\mu_{s}(p)\!-\!1$. 

As for the term $\ell\!=\!e$ in Eq.~(\ref{eq:3}) in which $\epsilon_{e}(p_{\rm F} 
\!+\! k)\!=\!v_{\rm F}k$, $\mu_{e}(p_{\rm F})\!=\!\mu_{\rm F}$, and 
$\phi_{e}(p_{\rm F})\!=\!\phi_{\rm F}$ with $\Omega_{e}(p_{\rm F})\!=\!
\Omega_{\rm F}$ for the spin-$\frac{1}{2}$ Luttinger model, we can easily see 
that $\partial A(p_{\rm F} \!+\! k,\omega) / \partial \omega$ diverges at 
$\omega \!=\! v_{\rm F}k$ for $0< \!\theta \!\leq 1/8$ even at $p \!\to \!
p_{\rm F}$. By using the identity $\int_{0}^{\infty}t^{\mu_{\rm F}}
{\rm e}^{\pm {\rm i}t}{\rm d} t \!=\! \Gamma ( \mu_{\rm F}\!+\!1)
{\rm e}^{\pm {\rm i} \pi (\mu_{\rm F}\!+\!1)/2}$, we can also verify that 
$A(p,\omega)$ behaves in accord with Eq.~(\ref{eq:4}), 
exhibiting a structure of the pseudoelectron. 
\begin{figure}[thbp]
\begin{center}
\includegraphics[scale=0.322,keepaspectratio]{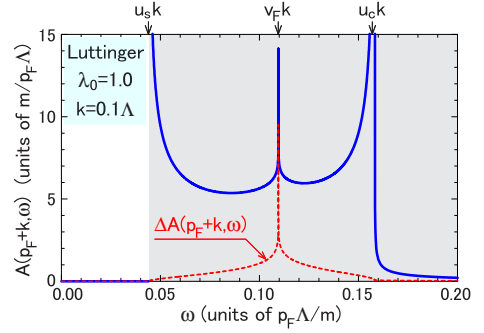}
\end{center}
\caption[Fig.1]{(Color online) Full spectral function $A(p_{\rm F}\!+\!k,\omega)$ 
at $k\!=\!0.1\Lambda$ for the spin-$\frac{1}{2}$ Luttinger model with use of 
$u_s\!=\! 0.4406 p_{\rm F}/m$, $u_c\!=\! 1.582 p_{\rm F}/m$, and $\theta \!=
\! 0.05351$ obtained by the Bethe ansatz for the Yang-Gaudin model with 
$\lambda_0 \!=\!1.0$ (solid curve).  
The contribution from a pseudoelectron excitation ($\omega \!=\! v_{\rm F} k$), 
which has been omitted in preceding studies~\cite{MS92,Voit93,OTS11}, is explicitly shown 
by $\Delta A(p_{\rm F} \!+\! k, \omega)$ (dotted curve).} 
\label{fig:1}
\end{figure}
In Fig.~\ref{fig:1}, we have explicitly shown the generic feature of 
$A(p, \omega)$ in which a peak with a cusp exists at the pseudoelectron mode 
[$\omega \!=\! \epsilon_{e}(p)$] in addition to the well-known double divergent 
peaks for $p$ near $p_{\rm F}$. In plotting $A(p, \omega)$, we need to know 
concrete values for $u_s$, $u_c$, and $\theta$; we have determined them by 
adopting the Yang-Gaudin model~\cite{Yang67} i.e. the 1D electron gas with a 
$\delta$-function interaction, described by the Hamiltonian $H \! =\!  
-\frac{1}{2m}\sum_{i=1}^{N} \frac{\partial^2}{\partial x_i^2}\! + \! V_0 
\sum_{i <j} \delta (x_i\!  -\!  x_j)$, in the intermediate-coupling regime 
of $\lambda_0 \! \equiv \! mV_0/2 \pi p_{\rm F}\!  = \! 1.0$.

Three comments are in order on the result in Fig.~\ref{fig:1}: (i) From its 
definition, $G(x,t)$, the inverse Fourier transform of $G(p,t)$, describes the 
space-time evolution of an additional electron injected to the system at 
$x\!=\!0$ and $t\!=\!0$. As long as $|x|$, $|v_{\rm F} t| \!\ll\! \Lambda^{-1}$, 
the injected electron is not much affected by the interactions and behaves like 
a free electron. This free-electron-like behavior has no effect on the leading 
singularities in $n(p)$, density of states, and $A(p, \omega)$, but it induces 
the subleading pseudoelectron singularity in $A(p, \omega)$. 
(ii) Irrespective of $\Lambda$, the limit of $\eta\!\to\!0^+$ is required in 
Eq.~(\ref{eq:1}) to keep the correct electron anticommutation relation 
$\{c_{p\sigma},c^{\dagger}_{p'\sigma'}\} \!=\! \delta_{pp'}\delta_{\sigma \sigma'}$ 
so as to satisfy the sum rule $\int_{-\infty}^{\infty} \! A(p,\omega){\rm d}\omega 
\!=\! \langle\{c_{p\sigma},c^{\dagger}_{p\sigma}\}\rangle \!=\!1$~\cite{Theumann77}. 
Thus the proper limiting procedure in Eq.~(\ref{eq:1}) is to make $\eta$ zero 
first with $\Lambda$ being finite, leading to the result in Fig.~\ref{fig:1}. 
If $\eta$ is replaced by $\Lambda^{-1}$ as was previously the case, the pole 
contribution disappears, because the residue vanishes with that choice of $\eta$, 
constituting the reason for the omission of the pseudoelectron contribution 
in preceding studies~\cite{MS92,Voit93,OTS11}. That choice is 
physically incorrect, because the anticommutation relation is not globally 
satisfied, for example, for $|x|\!\ll\!\Lambda^{-1}$, violating the sum rule.
(iii) In the course of obtaining $A(p,\omega)$, we need to deal with a double 
Fourier transform, which is best implemented by use of a mathematical trick 
virtually identical to Feynman parameters~\cite{Peskin, OTS11}. By introducing 
two Feynman parameters, we can analytically perform the double Fourier transform 
and then resort to numerical computation of the remaining double Feynman integral. 
The result thus computed with high accuracy is plotted in Fig.~\ref{fig:1} 
and exhibits a singular behavior precisely matched up with the one suggested 
in Eq.~(\ref{eq:4}). 

Now let us discuss the effect of nonlinear dispersion on $A(p,\omega)$ as well 
as the case of $|p|$ far away from $p_{\rm F}$ by adopting the mobile impurity 
model which was previously introduced for treating the spinon and holon modes 
without spoiling the rigor of the whole theory~\cite{IG09,SIG10}. Our strategy 
is to employ the same model to calculate the contribution to $A(p,\omega)$ from 
the pseudoelectron excitation. In the model, we first consider an {\it electron} 
with either linear or nonlinear dispersion $\xi_p$ as a mobile impurity in the 
LL with the velocity $v_p \!=\! \partial \xi_p/\partial p$. (Here we take only 
the case of $\xi_p \! \ge \! 0$, i.e., $|p| \! \geq \! p_{\rm F}$.) This impurity 
electron is assumed to couple with the spin and charge collective excitations 
near the right ($+$) and left ($-$) Fermi points with the interaction 
${\tilde V}_{\alpha \nu}\!\equiv\!{\tilde V}_{\alpha \nu}(p)$ ($\alpha \!=\! 
\pm$, $\nu \!=\! s$, $c$)~\cite{comment2}. Then, we can calculate the Green's 
function for the mobile impurity electron by straightforwardly extending the 
method developed by Dzyaloshinski\u i and Larkin~\cite{DL71,Metzner98} or in a 
more general GW$\Gamma$ framework~\cite{Takada01,MT11}, from which we obtain an 
integral equation to determine the Green's function $G(p\!+\!k,\omega)$ in 
momentum-frequency space as
\begin{align}
& \!\! \left( \omega  \!-\! \xi_p  \!-\! v_p k  \right) G(p\!+\!k, \omega) 
\nonumber
\\
&  \!=\! 1 \!+\! {\rm i} \! \int \frac{{\rm d} q {\rm d} \varepsilon}{(2 \pi)^2} 
\frac{W (q,\varepsilon)}{\varepsilon \! - \! v_p q} G(p \!+\! k \!-\!q, \omega 
\!- \! \varepsilon)  {\rm e}^{- |q|/\Lambda},
\label{eq:5}
\end{align}
with $\Lambda$ the cutoff of momentum transfer $q$ and $W(q,\varepsilon)\!=\!
(1/4) \sum_{\alpha \beta \nu}{\tilde V}_{\alpha \nu}
\chi_{\alpha \beta \nu}(q,\varepsilon){\tilde V}_{\beta \nu}$ with use of 
$\chi_{\alpha \beta \nu}(q,\varepsilon)$ the exact correlation functions 
for the Luttinger model. 

Since Eq.~(\ref{eq:5}) is very similar to the one known well in the Luttinger 
model~\cite{Solyom79}, we are familiar with its solution; by transforming the 
variables from momentum and frequency into real space and time, we can write 
down a differential equation for $G(x,t)$, from which we obtain
\begin{align}
& G (p\!+\!k,t) \!=\! 
\frac{ - {\rm i} {\rm e}^{- {\rm i} \epsilon_{e}(p + k) t}  }{
\displaystyle{
\prod_{{\nu} = {c}, {s}} \prod_{{\alpha} = \pm}\! \left[ 1\!+\!{\rm i} 
(u_{\nu} \!-\! \alpha v_p) \Lambda t \right]^{
\left(\delta_{\alpha \nu}/2\pi \right)^2/2} } } , 
\label{eq:6}
\end{align}
for $t> 0$, where $\epsilon_e(p\!+\!k)$ the dressed impurity energy contains 
an energy shift in proportion to the cutoff $\Lambda$ as
\begin{align}
\epsilon_e(p\!+\!k) = \xi_p \!+\! v_p k \!-\! \Lambda \!\sum_{\alpha \nu} 
\left( \delta_{\alpha\nu}/2\pi \right)^2 (u_\nu \!-\! \alpha v_p)/2. 
\label{eq:7}
\end{align}
In Eqs.~(\ref{eq:6}) and ({\ref{eq:7}), phase shifts $\delta_{\alpha \nu} 
\!\equiv\! \delta_{\alpha \nu}(p)$ are {\it independent} of $\Lambda$ and 
related to ${\tilde V}_{\alpha \nu}$ and $v_p$ through 
\begin{align}
\delta_{\pm c} \!=\! 
\frac{{\tilde V}_{\pm c} \sqrt{1 \!+\! \theta} \!-\! {\tilde V}_{\mp c} 
\sqrt{\theta} }{u_c \!\mp\! v_p}, 
\quad 
\delta_{\pm s} \!=\! \frac{{\tilde V}_{\pm s} }{u_s \!\mp\! v_p} .
\label{eq:8}
\end{align}
The results in Eqs.~(\ref{eq:7}) and~(\ref{eq:8}) indicate that 
$\epsilon_{e}(p_{\rm F} \!+\! k) \!=\! v_{\rm F}k$, $\delta_{+c}(p_{\rm F})\!=
\! 2 \pi \sqrt{1 + \theta}$, $\delta_{-c}(p_{\rm F}) \!=\! 2 \pi \sqrt{\theta}$, 
$\delta_{+s}(p_{\rm F}) \!=\! 2 \pi$, and $\delta_{-s}(p_{\rm F}) \!=\! 0$. 
Thus at $p \!=\! p_{\rm F}$, we see that Eq.~(\ref{eq:6}) is reduced to 
the pole contribution in Eq.~(\ref{eq:1}). 

\begin{figure}[thbp]
\begin{center}
\includegraphics[scale=0.305,keepaspectratio]{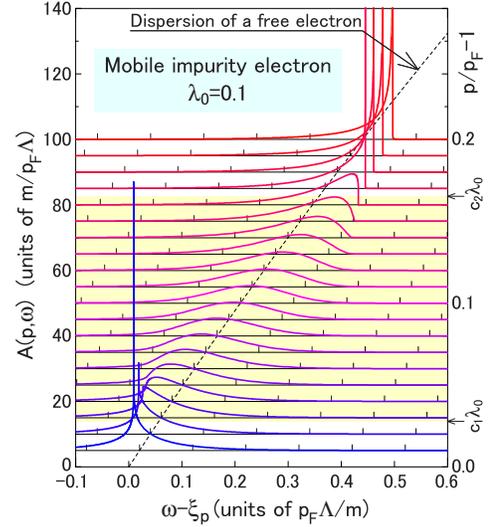}
\end{center}
\caption[Fig.2]{(Color online) Structural evolution of $A(p, \omega)$ as a 
function of $\omega$ in the very vicinity of $\xi_p$ with $p$ going away 
from $p_{\rm F}$  (from $p/p_{\rm F} -1 \!=\!0.01$ to $p/p_{\rm F} -1 \!=\! 0.2$ 
by the interval $0.01$) in the mobile impurity model at $\lambda_0 = 0.1$.}
\label{fig:2}
\end{figure}

For illustration of the overall behavior of $A(p \!>\! p_{\rm F},\omega)$ with the change 
of $p$ and $\omega$, we adopt the Yang-Gaudin model in the weak-coupling region 
($\lambda_0\!\leq\!0.1$) in which analytic expressions such as $v_{\rm F}\!=\!
p_{\rm F}/m$, $u_s \!=\!(1\!-\!\lambda_0)v_{\rm F}$, $u_c \!=\!\sqrt{1\!+\!2 
\lambda_0}v_{\rm F}$, and $\theta \!=\!(\sqrt{1\!+\!2\lambda_0}\!-\!1)^2/
4\sqrt{1\!+\!2\lambda_0}$ are successfully checked to reproduce the exact 
Bethe-ansatz results with sufficient accuracy, indicating that we may well 
specify the mobile impurity model by employing the weak-coupling results of the 
quadratic dispersion $\xi_p \!=\! (p^2\!-\!p_{\rm F}^2)/2m$ and the interactions, 
${\tilde V}_{\pm c}(p) \!=\! -{\tilde V}_{+s}(p)\!=\!V_0$, and ${\tilde V}_{-s} (p) \!
=\!V_0/(1\!-\!2\lambda_0 \ln [\xi_p/E_i])$, all of which are obtained by the 
poor man's scaling with $E_i$ an initial energy scale. Then, from $G(p,t)$ in 
Eq.~(\ref{eq:6}), we can explicitly compute $A(p,\omega)$ which is insensitive 
to the choice of $E_i$. In Fig.~\ref{fig:2}, $A(p,\omega)$ thus obtained at 
$\lambda_0\!=\!0.1$ is displayed with increasing $p$ from $p_{\rm F}$ to show 
its complete structural evolution in the 1D weakly-interacting electron gas 
with quadratic dispersion. Since we focus on the region of $\omega$ in the very 
vicinity of $\xi_p$, only the pseudoelectron mode appears as a singular 
structure in $A(p,\omega)$ in Fig.~\ref{fig:2}. 

\begin{figure}[bthp]
\begin{center}
\includegraphics[scale=0.271,keepaspectratio]{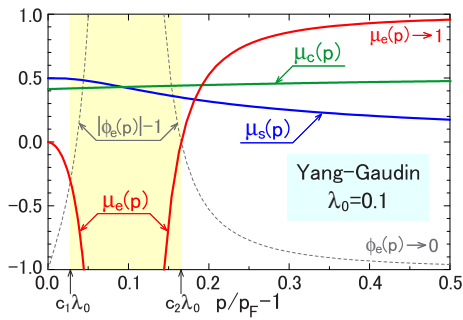}
\end{center}
\caption[Fig.3]{(Color online) Leading exponents of the spectral function at 
the spinon, antiholon, and pseudoelectron modes, which are denoted 
by $\mu_{s}(p)$, $\mu_{c}(p)$, and $\mu_{e}(p)$, respectively, versus 
$p/p_{\rm F} - 1$ for the Yang-Gaudin model with $\lambda_0 = 0.1$.}
\label{fig:3}
\end{figure}

By comparing Eq.~(\ref{eq:6}) in its long-time limit with Eq.~(\ref{eq:3}), 
we can determine $\mu_{e}(p)$ and $\phi_{e}(p)$ as
\begin{align}
&\mu_e(p) = 1 - \sum_{\alpha \nu} \left( \delta_{\alpha\nu}/2\pi \right)^2/2  , 
\label{eq:9}
\\
&\phi_e(p) = -\sum_{\alpha \nu} \left( \delta_{\alpha\nu}/2\pi \right)^2 
{\rm sign}(u_\nu - \alpha v_p)/2. 
\label{eq:10}
\end{align} 
They are {\it independent} of $\Lambda$ and universal, 
but they depend on $p$ in the system with nonlinear dispersion~\cite{Meden99}, 
as plotted in Fig.~\ref{fig:3} in which $\mu_e(p)$ and $|\phi_e(p)|\!-\!1$ are 
given by the solid and dotted curves, respectively. The structural feature of 
$A(p,\omega)$ is specified by both $\mu_e(p)$ itself and its relative value with 
respect to $|\phi_e(p)| \!-\! 1$. Thus we define two dimensionless constants, 
$c_1 \!=\! 1 \!-\!1/\sqrt{2} \!+\! O(\lambda_0)$ and $c_2\!=\!\sqrt{3} \!+\! O(\lambda_0)$, 
corresponding to the intersections of the solid and dotted curves. By virtue of 
Eq.~(\ref{eq:4}), $A(p,\omega)$ at $\omega\!=\!\epsilon_{e}(p)$ has a peak with 
a cusp for $p/p_{\rm F} \!-\! 1 \!<\! c_1 \lambda_0$ (the spin-charge separated 
regime) and a divergent one for $p/p_{\rm F} \!-\! 1 \!>\! c_2 \lambda_0$ except 
for the narrow vicinity of $p/p_{\rm F} \!-\! 1 \!=\! c_2 \lambda_0$ (the nearly 
free-electron regime). In the intermediate region of 
$c_1\lambda_0 \!<\! p/p_{\rm F} \!-\! 1 \!<\! c_2\lambda_0$, on the other hand, 
$A(p,\omega)$ shows only a broad peak around $\omega\!=\!\xi_p$ due to the 
strong damping effect brought about by the allowed energy-conserving 
electron-charge excitation scatterings in which $\xi_p\!-\!\xi_{p-q}\!=\!u_c q$ 
is satisfied in Eq.~(\ref{eq:5}) (the pseudoelectron damping regime).

In order to give an entire picture of $A(p,\omega)$ including spinon and antiholon 
modes, we have also plotted the exponents, $\mu_{s}(p)$ and $\mu_{c}(p)$, 
in Fig.~\ref{fig:3}, which are exact for the Yang-Gaudin model with $\lambda_0
\!=\! 0.1$. (Accurate values for those exponents can be derived from 
the exact Bethe-ansatz results for $\epsilon_{s}(p)$ and $\epsilon_{c}(p)$ for 
Galilean invariant systems~\cite{SIG10}.) By comparing them with $\mu_{e}(p)$, 
we can identify the mode(s) to dominate $A(p,\omega)$: (i) For $|v_p|/v_{\rm F} 
\!-\! 1 \!=\! |p|/p_{\rm F}\!-\!1\! \ll \! \lambda_0$, the situation is characterized 
by $\mu_{s}(p)\!\approx\!\mu_{c}(p)\!\gg\!\mu_e(p)$, implying the dominance of 
spinon and antiholon modes. Thus the physics in this regime is well described 
by the concept of spin-charge separation, but the pseudoelectron is now found 
to appear as an additional singular cusp structure as depicted in Fig.~\ref{fig:1}}. 
(ii) For $|v_p|/v_{\rm F} \!-\! 1 \!\approx\! \lambda_0$, the pseudoelectron 
excitation is overdamped as shown in Fig.~\ref{fig:2} and sandwiched between 
the spinon and antiholon divergent peaks, making it difficult to be detected, 
although we do not expect that its total contribution to the spectral weight is 
negligible. 
(iii) For $|v_p|/v_{\rm F} \!-\! 1 \!\gg\! \lambda_0$, $\mu_{e}(p)$ becomes much 
larger than either $\mu_{s}(p)$ or $\mu_{c}(p)$, indicating that the long-time 
evolution of $G(p,t)$ is controlled by the single mode of pseudoelectron, 
well defining the nearly free-electron regime. In fact, according to our 
explicit calculation based on Ref.~\cite{SIG10}, $\mu_s(p)$ is a decreasing 
function of $|p|$ and becomes negative for $|p|\! \agt \!2p_{\rm F}$, so that 
the edge of support of $A(p,\omega)$ never diverges. As for the antiholon, 
$\mu_c(p)$ slowly increases with increasing $|p|$ and reaches $1/2$ at $|p| 
\!\to\! \infty$. In the large-$|p|$ limit with $\lambda_0$ fixed, both 
$\epsilon_{e}(p)$ and $\epsilon_{c}(p)$ approach the free-electron dispersion, 
but because $\mu_{e}(p)\!\approx\!2\mu_{c}(p)\!\approx\!1$, the antiholon peak 
gets absorbed into the pseudoelectron one~\cite{comment3}. Thus $A(p,\omega)$ is 
composed of a single divergent peak, reduced to a delta function at $\omega \!=\! 
\xi_p$ at $|p|\! \to \! \infty$ where $G(p,t)\!=\! -{\rm i}{\rm e}^{- {\rm i} 
\xi_p t}$ due to $\mu_{e}(p) \!\to\! 1$ and $\phi_{e}(p) \!\to\! 0$ in Eq.~(\ref{eq:3}).

Four comments are in order: 
(i) As is evident from Eqs.~(\ref{eq:8})--(\ref{eq:10}), both $\mu_e(p) $ and 
$\phi_e(p)$ depend on $p$ mainly through $v_p\!=\!\partial \xi_p/\partial p$. 
For non-Galilean invariant systems, we can obtain very similar results for them 
by suitably choosing the dispersion $\xi_p$ and an interaction parameter 
$\lambda_0$, at least for $|v_p| \!\geq\! v_{\rm F}$ in the weak-coupling 
region~\cite{comment4}. 
(ii) The strong pseudoelectron damping occurs on the condition $|v_p|\!\approx\!
u_c$, which is allowed in general at $|v_p|/v_{\rm F} \!-\! 1\!\sim\!
{\rm min}(1,\lambda_0)$. Thus, if $\lambda_0$ lies outside of the weak-coupling 
region, the spin-charge separated regime appears for $|v_p|/v_{\rm F} \!-\! 1 
\!\ll\!{\rm min}(1,\lambda_0)$. 
(iii) Our discussion has been limited only for $|p| \!\geq\! p_{\rm F}$, but a 
similar discussion can be made for $|p| \!<\! p_{\rm F}$ in which electron-hole 
asymmetry due to nonlinearity in $\xi_p$ may be a matter of interest. 
(iv) According to the numerical results for $A(p,\omega \!<\! 0)$ obtained by 
the dynamical density-matrix renormalization group (DDMRG) method for the 1D 
Hubbard model in the intermediate coupling regime of $U\!=\!4.9t$ and the filling 
$n\!=\!0.6$ (for which $\lambda_0\!\sim\!1$)~\cite{BGJ04}, the peaks of spinon, 
holon, and its shadow band for $|p| \!<\! p_{\rm F}$, which are predicted by the 
Bethe ansatz, are found, but there appears no signature of an additional noticeable 
structure, at least for $|p|$ not close to $p_{\rm F}$~\cite{comment5}. This 
absence of the pseudoelectron structure in the DDMRG does not contradict our 
theory, because the data without a strong artificial broadening effect are 
given only in the pseudoelectron damping regime of $1\!-\! |v_p|/v_{\rm F} 
\!\sim\!\lambda_0$. 

In summary, we have theoretically studied the overall behavior of $A(p,\omega)$ 
in a 1D metal at zero temperature in order to establish the concept of 
a ``pseudoelectron'', describing the behavior of an electron injected into the 
LL with either linear or nonlinear dispersion. This pseudoelectron 
is found to manifest itself in an entire range of $p$, though its importance in 
the whole structure of $A(p,\omega)$ depends on $p$; for $|p|\!\approx\!p_{\rm F}$, 
it appears only as an additional cusp structure to the main peaks of spinon and 
(anti)holon, while for $|p|\! \gg \!p_{\rm F}$, it provides a main and divergent peak. 
This pseudoelectron very much resembles a quasiparticle 
in higher-dimensional Fermi-liquid systems (e.g., see Fig.~3 in Ref.~\cite{MT11}), 
although it is not quite the same, reflecting the specialty of 1D physics. 
We hope that this concept of a pesudoelectron will be confirmed in the future 
through experiment and/or large-scale numerical calculation with 
deliberately-chosen parameters so as to avoid its overdamping regime. 

H.M. thanks H. Fukuyama, M. Nakamura, and K. Penc for valuable comments. 
This work is partially supported by Innovative Area "Materials Design through 
Computics: Complex Correlation and Non-Equilibrium Dynamics" (No. 22104011) 
from MEXT, Japan. 



\begin{references}

\bibitem{Tomonaga50} S. Tomonaga, Prog. Theor. Phys. {\bf 5}, 544 (1950); 
J. M. Luttinger, J. Math. Phys. {\bf 4}, 1154 (1963); 
F. D. M. Haldane, J. Phys. C {\bf 14}, 2585 (1981). 

\bibitem{Experiment} See, for example, Y. Jompol {\it et al}., Science {\bf 325}, 
597 (2009); V. V. Deshpande {\it et al}., Nature {\bf 464}, 209 (2010).

\bibitem{ISG12} For a review, see A. Imambekov, T. L. Schmidt, and L. I. Glazman, 
Rev. Mod. Phys. {\bf 84}, 1253 (2012). 

\bibitem{Penc95} K. Penc, F. Mila, and H. Shiba, Phys. Rev. Lett. {\bf 75}, 
894 (1995); K. Penc {\it et al}., Phys. Rev. Lett. {\bf 77}, 1390 (1996); 
Phys. Rev. B {\bf 55}, 15475 (1997); 
J. M. P. Carmelo {\it et al}., J. Phys.: Condens. Matter {\bf 18}, 5191 (2006).

\bibitem{IG09} A. Imambekov and L. I. Glazman, Science {\bf 323}, 228 (2009); 
Phys. Rev. Lett. {\bf 102}, 126405 (2009).

\bibitem{PWA09} R. G. Pereira, S. R. White, and I. Affleck, 
Phys. Rev. B {\bf 79}, 165113 (2009).

\bibitem{SIG10} T. L. Schmidt, A. Imambekov, and L. I. Glazman, 
Phys. Rev. Lett. {\bf 104}, 116403 (2010); Phys. Rev. B {\bf 82}, 245104 (2010).

\bibitem{Essler10} F. H. L. Essler, Phys. Rev. B {\bf 81}, 205120 (2010). 

\bibitem{Meden99} V. Meden, Phys. Rev. B {\bf 60}, 4571 (1999); Meden claims that 
$\mu_{\nu}(p)$ for $|p| \!\neq\! p_{\rm F}$ in the LL with strictly linear 
dispersion depends on the ultraviolet regularization. We can easily solve this 
difficulty by considering a nonlinear dispersion~\cite{IG09,PWA09,SIG10,Essler10}, 
leading to physically meaningful $\mu_{\nu}(p)$.

\bibitem{Solyom79} J. S\'{o}lyom, Adv. Phys. {\bf 28}, 201 (1979). 

\bibitem{Suzumura80} Y. Suzumura, Prog. Theor. Phys. {\bf 63}, 51 (1980).

\bibitem{comment1} In standard notations~\cite{Solyom79}, we have taken 
$g_{4\parallel} \!=\! 0$ and $g_{2\parallel} \!=\! g_{2\perp}$ by the 
Pauli principle and spin-SU(2) symmetry, respectively, and defined 
$g_4 \!\equiv\! g_{4\perp}$ and $g_2 \!\equiv\! g_{2\parallel} \!=\! g_{2\perp}$.

\bibitem{SM93} K. Sch\"{o}nhammer and V. Meden, Phys. Rev. B {\bf 47}, 
16205 (1993); {\it ibid}. {\bf 48}, 11390 (1993).

\bibitem{Yang67} C. N. Yang, Phys. Rev. Lett. {\bf 19}, 1312 (1967); 
M. Gaudin, Phys. Lett. A {\bf 24}, 55 (1967).

\bibitem{Theumann77} A. Theumann, Phys. Rev. B {\bf 15}, 4524 (1977). 

\bibitem{MS92} V. Meden and K. Sch\"{o}nhammer, Phys. Rev. B {\bf 46}, 
15753 (1992); J. Voit, Phys. Rev. B {\bf 47}, 6740 (1993).

\bibitem{Voit93} J. Voit, J. Phys.: Condens. Matter {\bf 5}, 8305 (1993); 
Voit calculated $A(p,\omega)$ without replacing $\eta$ by $\Lambda^{-1}$ 
for two special cases of the spinless ($g_{2\parallel}\!\neq\!0$, $g_{2\perp}\!
=\!g_{4\perp}\!=\!0$) and the one-branch ($g_{4\perp}\!\neq\!0$, $g_{2\parallel}
\!=\!g_{2\perp}\!=\!0$) Luttinger models. In the former case, the pole 
contribution leads to not a peak-like cusp but a {\it shoulder} at $\omega \!=
\! v_{\rm F} k$ in such a way as $A(p_{\rm F} \!+\! k,\omega) \!\propto
 {\rm const.} \!-\! |\omega \!-\! v_{\rm F} k|^{1\!+\theta}/(\omega \!-\! 
v_{\rm F} k)$, while for the latter, $A(p,\omega)$ {\it diverges} 
logarithmically at $\omega \!=\! v_{\rm F} k$, causing a {\it finite} jump 
in $n(p)$, a pathological nature of the model. Voit guessed some electron-like 
structure in $A(p,\omega)$ from those results. In a sense, his guess is now 
substantiated by us and culminated in a generic pseudoelectron-excitation 
structure in the spin-$\frac{1}{2}$ LL. 

\bibitem{OTS11} E. Orignac, M. Tsuchiizu, and Y. Suzumura, 
Phys. Rev. B {\bf 84}, 165128 (2011).

\bibitem{Peskin} M. E. Peskin and D. V. Schroeder, {\it An Introduction to 
Quantum Field Theory} (Addison-Wesley, New York, 1995).
\bibitem{comment2} In the LL with linear dispersion, $\xi_{p_{\rm F}+k}
\!=\! v_{\rm F} k$, ${\tilde V}_{+c}(p_{\rm F}) \!=\! g_4$, 
${\tilde V}_{-c}(p_{\rm F})\!=\!2 g_2$, ${\tilde V}_{+s}(p_{\rm F})\!=\!-g_4$, 
and ${\tilde V}_{-s}(p_{\rm F}) \!=\! 0$. 

\bibitem{DL71} I. E. Dzyaloshinski\u i and A. I. Larkin, Zh. Eksp. Teor. Fiz. 
{\bf 65}, 411 (1973) [Sov. Phys. JETP {\bf 38}, 202 (1974)].

\bibitem{Metzner98} For extension to higher dimensions, see 
W. Metzner, C. Castellani, and C. D. Castro, Adv. Phys. {\bf 47}, 317 (1998). 

\bibitem{Takada01} Y. Takada, Phys. Rev. Lett. {\bf 87}, 226402 (2001). 

\bibitem{MT11} H. Maebashi and Y. Takada, Phys. Rev. B {\bf 84}, 245134 (2011). 

\bibitem{comment3} In our notation, $\mu_c(p)$ is the exponent along the main 
branch of the antiholon spectrum. There also exist its shadow branches, but 
all the exponents associated with these shadow branches become negative for 
$|p| \to \infty$ as long as $\lambda_0$ is finite. At $\lambda_0 \!\to\! \infty$, 
the largest one among them becomes equal to $\mu_c(p)$ with their actual value 
being $3/8$ as derived in the Hubbard model by \cite{Penc95}.

\bibitem{comment4} In the 1D Hubbard model outside of the weak-coupling region 
(for which $\lambda_0\! \agt \!1$), there is no nearly free-electron regime of 
$|\! \sin p|/\sin p_{\rm F} \!-\! 1 \!\gg\! \lambda_0$ unless the filling 
$n \!\ll\! 1$. 

\bibitem{BGJ04} H. Benthien, F. Gebhard, and E. Jeckelmann, 
Phys. Rev. Lett. {\bf 92}, 256401 (2004).

\bibitem{comment5} The data for $\mu_s(p)$ in \cite{BGJ04} are in good 
accord with those obtained in \cite{Essler10} and consistent with the results 
in other works \cite{Penc95,SIG10}. 
It is, however, noted that the data of $\mu_{c}(0) \!=\! \mu_{c'}(0)\!=\!0.70$ 
and $\mu_{c}(\pi/10) \!=\! 0.44 \!<\! \mu_{c'}(\pi/10) = 0.56$ in \cite{BGJ04} 
with $\mu_{c'}(p)$ being the the shadow-band exponent 
contradict the results in those works~\cite{Penc95,SIG10}, 
where $\mu_c(0) \!=\! \mu_{c'}(0) \!\approx\! 3/8$ for $\lambda_0\! \agt \!1$ 
and $\mu_c(p) \!\geq\! \mu_{c'}(p)$ for any $p$.

\end{references}
\end{document}